\documentclass[12pt]{article}
\usepackage{a4wide}
\usepackage{graphics}
\usepackage{amsfonts}
\usepackage{amstext}
\usepackage{amsmath}
\usepackage{amssymb}
\usepackage{amsthm}
\usepackage{psfrag}
\usepackage{cite}

\def\Ai{{\rm Ai}}

\def\G{\Gamma}
\def\K{\mathcal{K}}
\def\vp{\varphi}
\def\t{\tau}

\begin{document}
\title{Spatial correlations of the 1D KPZ surface on a flat substrate}
\author{
\vspace{5mm}
T. Sasamoto
{\footnote {\tt e-mail: sasamoto@stat.phys.titech.ac.jp}}
\\
{\it Department of Physics, Tokyo Institute of Technology,}\\
\vspace{5mm}
{\it Oh-okayama 2-12-1, Meguro-ku, Tokyo 152-8551, Japan}\\
}
\maketitle

\begin{abstract}
We study the spatial correlations of the one-dimensional 
KPZ surface for the flat initial condition. 
It is shown that the multi-point joint distribution for
the height is given by a Fredholm determinant, 
with its kernel in the scaling limit explicitly obtained. 
This may also describe the dynamics of the largest
eigenvalue in the GOE Dyson's Brownian motion model.
Our analysis is based on a reformulation of the 
determinantal Green's function for the totally ASEP
in terms of a vicious walk problem.
\end{abstract}

Surface growth has been an important subject of physics 
both from practical and fundamental aspects. 
While a good control of it is crucial in recent 
atom-scale technology, a rich variety of 
interesting surface pattern has
attracted much attention of theoretical studies
\cite{Barabasi1995,Me1998}. 
It is also important from the point of view of 
noneqilibrium statistical mechanics.

It is in general difficult to obtain detailed information 
about the properties of surface by analytical methods. 
However in one spatial dimension some surface growth 
models are known to be exactly solvable. They are 
very special in many respects, but give us a lot of 
insight to understand the properties of surfaces in nature. 
The Kardar-Parisi-Zhang (KPZ) equation, introduced in 
\cite{KPZ1986}, is one of the minimal models in the 
theory of surface growth which have both nonlinear 
and noise effects. A lot of models were shown to 
belong to the same universality class as the KPZ 
equation, i.e., the KPZ universality class \cite{KS1992}.
But the analysis had been mainly restricted to the 
exponents for some time. 

A next breakthrough comes from an observation that some 
surface growth models, in particular the polynuclear 
growth (PNG) model, are related to combinatorial 
problem of Young tableaux \cite{BDJ1999}.
In \cite{Jo2000,PS2000a}, the height fluctuation of the 
surface in the KPZ universality class was shown 
to be equivalent to that of the largest eigenvalue of 
random matrices \cite{Mehta2004}. 
One of the important results is that the height 
fluctuation at one point strongly depends on boundary 
and initial conditions \cite{BR2000,PS2000a,PS2002p}.
For instance, droplet growth, in which the surface 
starts from a seed and grows into a droplet shape, is
related to the Gaussian unitary ensemble (GUE); the 
model on a flat substrate is related to the Gaussian 
orthogonal ensemble (GOE).

More recently spatial correlations of the surface, 
which has an information about how rough the surface is,
have been calculated. 
For the droplet growth, it was shown to be equivalent to 
the dynamics of the largest eigenvalue of the 
GUE Dyson's Brownian motion model 
\cite{Dyson1962,PS2002b,Johansson2003,Johansson2005}. 
The PNG model with external sources and that in half-space
have also been studied \cite{SI2004,IS2004b,BR2004p}.
All these results have been obtained by introducing the 
multi-layer version of the PNG model 
\cite{PS2002b,Johansson2003}. 

The spatial correlations for the flat case is important 
for several reasons. This case is suitable for studying
how an initially straight surface without fluctuation grows 
into a rough one. A lot of simulation have been performed 
for the flat case.
In addition, there is a more theoretical interest.
It is possible to define the multi-layer PNG model 
for the flat case and the fluctuation of the 
multi-layers at one point are shown to be the same as that 
of the largest eigenvalues of the GOE \cite{Ferrari2004}.
Combined with the spatial homogeneity of the flat case,
this suggests that the flat case is related to the GOE 
Dyson's Brownian motion model\cite{Dyson1962}, 
which is equivalent to the $\beta=1$ case of the Calogero 
model \cite{Calogero1969,Sutherland1972}.
Since this is not a free fermion model, the analysis 
of the flat case might well be qualitatively difficult 
than for other cases. 
Note that the constraints of the multi-layers are complicated 
and have not allowed us to study the multi-point fluctuation. 

In this article we tackle this problem by a related 
but a different method, i.e., by utilizing a formula
for the one-dimensional totally asymmetric simple 
exclusion process (TASEP) on infinite lattice.
There are several versions of the update rules for TASEP.
Here we consider the continuous time version, 
which is most well studied. 
In TASEP, each site is either occupied by a particle 
or empty. In an infinitesimal time duration $dt$ each particle 
tries to hop to the right nearest site with probability $dt$. 
The hopping is not allowed if the target site 
is already occupied by another particle. 
This is a very simple model but exhibits a lot of 
interesting behaviors 
\cite{Liggett1985,Liggett1999,Schuetz2000,Spohn1991}. 

We can interpret the TASEP dynamics as a kind of 
surface growth model if we replace each occupied site 
with a slope with -45 degree and each 
empty site with a slope with 45 degree. 
See Fig. 1. A hopping of a particle corresponds to 
a local surface growth in the surface growth picture. 
This surface growth model is known to belong to the 
KPZ universality class \cite{KS1992}. 
The flat initial condition corresponds to the initial 
surface configuration in which $\diagup$ and $\diagdown$
occurs alternately. 
We study the spatial correlations of this model.

If we describe the stochastic dynamics of TASEP 
in terms of a master equation, it turns out that
the transition rate matrix can be written 
as a kind of non-hermitian spin chain 
\cite{GS1992,Schuetz2000}.
In the language of spin chains, the TASEP is a kind of 
XXZ model and hence is {\it not} a free fermion model. 
If one applies the Bethe ansatz method to TASEP, 
the S-matrix is not just -1. But the Bethe ansatz analysis
is useful to understand the temporal properties of the 
ASEP \cite{GS1992,DL1998}. For instance the Bethe ansatz method 
allows us to construct the Green's function in the form of 
determinant \cite{Schuetz1997}.
Let $G(x_1,x_2,\cdots,x_N;t|y_1,y_2,\cdots,y_N;0)$
denote the probability that the $N$ particles are on sites 
$x_1,x_2,\cdots,x_N$ ($x_N < \cdots <x_1$) at time $t$ under 
the condition that they are on sites $y_1,y_2,\cdots,y_N$\,
($y_N < \cdots < y_1$) at time $0$.
Then $G(x_1,x_2,\cdots,x_N;t|y_1,y_2,\cdots,y_N;0)$ 
is given by 
\begin{align}
G(x_1,x_2,\cdots,x_N;t|y_1,y_2,\cdots,y_N;0) 
=
\det[F_{k-j}(x_{N-k+1}-y_{N-j+1};t)]_{j,k=1,\cdots,N}.
\end{align} 
Here the function $F_n(x;t)$ is defined by
\begin{equation}
 F_n(x;t) 
 =
 e^{-t}\frac{t^x}{x!}\sum_{k=0}^{\infty} (-1)^k \frac{(n)_k}{(x+1)_k}
 \frac{t^k}{k!}.
\end{equation}
When we fix $y_j$'s, the initial configuration of particles, 
we call this quantity $G(x_1,x_2,\cdots,x_N;t)$ as well.
This formula has already been used for studying the 
fluctuation properties of the TASEP and a discrete 
version of it \cite{NS2004,RakosSchuetz2004p}.

Now, on the other hand, let us consider a weight 
in the form of products of determinants,
\begin{equation}
 \prod_{r=1}^{N-1}
 \det[\phi(x_j^r,x_k^{r+1})]_{j,k=1}^{r+1}
 \det[\psi_j^{(N)}(x_{k+1}^N)]_{j,k=0}^{N-1},
\label{weight}
\end{equation}
on $x_j^r$ ($r=1,\cdots,N,\,j=1,\cdots,r$).
We put the condition $x_1^2 < x_2^2,x_1^3<x_2^3<x_3^3,\cdots$ and 
the convention $\phi(x_{r+1}^r,x_k^{r+1})=1$ for 
$r=1,\cdots,N-1, k=1,\cdots,r+1$.
Here the functions $\psi_j^{(r)}(x)$ and $\phi(x,y)$ 
are defined by 
\begin{align}
 \psi_j^{(r)}(x)    
 = 
 (-1)^{r-1-j} F_{-r+1+j}(x-y_{j+1},t) 
\end{align}
($j=0,\cdots,N-1$) and $\phi(x,y) = 0 \,(x_1>x_2), -1\,(x_1 \leq x_2)$.
Let $\mathbb{P}$ denote the corresponding measure. Then 
we have
\begin{align}
 G(x_1,\cdots,x_N;t) 
 =
 \mathbb{P}[x_1^r=x_r \,(r=1,2,\cdots,N)].
\label{GP}
\end{align}
This is one of the main results of this letter and 
is proved as follows. First one shows
\begin{align}
 \prod_{r=1}^{N-1} \sum_{x_{r+1}^{r+1}>\cdots>x_2^{r+1}(>x_1^{r+1})}
 \prod_{r=1}^{N-1}
 \det[\phi(x_j^r,x_k^{r+1})]_{j,k=1}^{r+1} f
 =
 \prod_{r=1}^{N-1} \prod_{l=2}^{r+1} \sum_{x_l^{r+1}=x_{l-1}^r}^{\infty} f.
\end{align}
with $f$ being an arbitrary antisymmetric function of $x_j^N$.
Here the summation on the left side is over all possible 
configurations of $x_j^r$'s, ($r=1,\cdots,N,j=2,\cdots,r$) with 
$x_1^r$ being fixed to $x_r$ ($r=1,\cdots,N$). 
This can be shown by considering when the determinant vanishes,
using the antisymmetry of the function $f$ and the mathematical 
induction. Then 
using a property of the function $F_n(x;t)$,
$F_{n+1}(x;t) = \sum_{y=x}^{\infty} F_n(y;t)$,
one arrives at (\ref{GP}).

If we interpret $r$ as a time coordinate and $x_j^r$ as the 
position of the $j$th walker at $r$, the weight in (\ref{weight})
can be interpreted as a kind of vicious walk problem with a 
peculiar structure that each particle is added at 
each time step. The formula (\ref{GP}) allows us to obtain the 
multi-point information of our surface model from the trajectory 
of the first particle, $x_1^r$ ($r=1,2,\cdots$), in the vicious 
walk problem. Vicious walk problem was introduced in \cite{Fisher1984} 
and is regarded as a sort of free fermion model \cite{Forrester1989}.
Hence (\ref{GP}) explicitly says that one can study the dynamics
of the TASEP in terms of free fermion even though the 
transition rate matrix does not look like a free fermion model. 

For each $r$ let us define another set of function, 
$\vp_j^{(r)}(x)$ ($j=0,\cdots,r-1$), which is a polynomial of 
order $j$, by the condition that they are orthonormal to 
$\psi_j^{(r)}(x)$'s,
\begin{equation}
\label{ortho}
 \sum_{x=-2r+2}^{\infty} \vp_j^{(r)}(x) \psi_k^{(r)}(x) = \delta_{jk}.
\end{equation}
We also define $ \phi_{r_1,r_2}(x_1,x_2)$ by 
\begin{align}
 \phi_{r_1,r_2}(x_1,x_2)
 &=
 (-1)^{r_2-r_1}\int_{\G_0}\frac{dz}{2\pi i}
 \frac{z^{x_1-x_2-1}}{(1-z)^{r_2-r_1}}
\end{align}
for $r_1<r_2$ and $\phi_{r_1,r_2}(x_1,x_2)=0$ for $r_1 \geq r_2$.
Notice $\phi_{r,r+1}(x_1,x_2) = \phi(x_1,x_2)$.

We can show that our vicious walk problem is determinantal,
i.e., the correlation functions of the system is written in 
the determinant form. For instance the probability that 
$x_l^{r_j}$ ($l=1,2,\cdots,r_j$) are occupied by walkers at $r_j$
($j=1,2$) is proportional to
\begin{equation}
\det[K(r_j,x_l^{r_j};r_k,x_m^{r_k})]
_{j,k=1,2,\, l=1,\cdots,r_j,\, m=1,\cdots,r_k}. 
\end{equation}
Here the matrix element is given by
\begin{align}
\label{K}
 K(r_1,x_1;r_2,x_2) 
 &=
 \tilde{K}(r_1,x_1;r_2,x_2) -\phi_{r_1,r_2}(x_1,x_2) \notag\\
 &=
 \sum_{j=0}^{N-1} \psi_j^{(r_1)}(x_1) \varphi_j^{(r_2)}(x_2)
 -\phi_{r_1,r_2}(x_1,x_2).
\end{align}
We set $\vp_j^{(r)}(x) =0$ for $j\geq r$ and hence the summation is 
actually up to $r_2-1$. 
This fact can be proved by following the strategy of \cite{NF1998}
but there appears some new feature compared to the usual case of 
fixed number of walkers \cite{BR2004p}. For instance, 
when $r_1<r_2$, $\psi_j^{(r)}$ with $j\geq r$'s are included in the 
summation of (\ref{K}).

Then the joint distribution of the first particle in our vicious 
walk problem, which corresponds to the joint distribution of 
the height in our original surface growth model, can be 
described by a Fredholm determinant with the kernel given by 
the function $K(r_1,x_1;r_2,x_2)$:
\begin{equation}
 \mathbb{P}[x_1^{r_j} > X_j \,(j=1,\cdots,m)] 
 =
 \det(1-K).
\end{equation}

Up to here our discussions are for general initial configuration.
We believe that there are many applications of the formula (\ref{GP})
to study the temporal properties of TASEP. In this article
we concentrate on analyzing the spatial correlations 
of the KPZ surface for the flat initial condition, which 
has not been solved by usual multi-layer PNG techniques. 

We take a special initial condition, 
$y_{j+1} = -2j (j=0,\cdots,N-1)$. See Fig. 1.
This is not exactly the flat initial condition,
but deep inside the negative ($x<0$) region, the 
correlations are the same as the flat case because 
the effect of the boundary is restricted near the 
origin at finite time.
One can also restrict the number of particles to $N$
since the dynamics of a particle can not affect the 
dynamics of particles on its right. 
Let $h(x,t)$ denote the surface height at position $x$
and at time $t$. The limiting shape is known to be
$h(x,t)/t \sim \frac12~(x\leq 0),
 ~\frac12\left(1+\frac{x^2}{t^2}\right)~(0< x \leq t),~
 x/t~(x\geq t)$.
We are interested in the fluctuation around this.

For the special choice $y_{j+1}=-2j$, 
one can find an explicit formula for $\vp_j^{(r)}(x)$,
\begin{equation}
 \vp_j^{(r)}(x)
 =
 \frac{(-1)^{r-1-j}}{2\pi i}\int_{\G_0}dz \frac{1+2z}{1+z}
 \frac{(1+z)^{x+r+j-1}}{z^{r-j}}e^{-zt},
\end{equation}
where $\G_0$ is a contour enclosing the origin anticlockwise.
It is not difficult to check the orthonormality relation (\ref{ortho}).
As a consequence the kernel has a double contour integral expression,
\begin{equation}
K(r_1,x_1;r_2,x_2) 
 =
 \int_{\G_{-1}}\frac{dw}{2\pi i} \int_{\G_0}\frac{dz}{2\pi i}
 \frac{(1+z)^{x_2+r_2-2}(-w)^{r_1}(1+2z)e^{(w-z)t}}
      {(1+w)^{x_1+r_1-1}(-z)^{r_2}(w-z)(1+w+z)} 
 -\tilde{\phi}_{r_1,r_2}(x_1,x_2)
\label{Kint}
\end{equation}
where
\begin{equation}
 \tilde{\phi}(x_1,x_2) 
 =
 \int_{\G_0}\frac{dz}{2\pi i} z^{x_2+r_2-x_1-r_1-1}(1-z)^{r_1-r_2}.
\end{equation}

Now we consider the scaling limit, in which universal properties 
of the model are expected to appear. 
First let us consider the positive ($x>0$) region. In this case,
it is convenient to set
\begin{align}
 r_j &= t \rho_j = t \rho + 2\sqrt{\rho} t^{2/3} \t_j/d, \\ 
 x_j + r_j -1 &= (1-\sqrt{\rho_j})^2t -(1-\sqrt{\rho}) d~ t^{1/3} \xi_j,
\end{align}
where $d = \rho^{-1/6}(1-\sqrt{\rho})^{-1/3}$ for $j=1,2$,
and take the $t\to \infty$ limit with $\rho$ fixed. 
Note that $0<\rho<1/4$ corresponds to looking at the positive 
($x>0$) region. Applying the saddle point analysis to (\ref{Kint}),  
the limiting kernel, $\K_2$, turns out to be that 
for the Airy process \cite{PS2002b,Johansson2003}.
Hence we conclude that the fluctuation of surface in the 
positive ($x>0$) region is the same as that of the droplet growth.
The spatial correlation of the surface is the same as the dynamics of 
the largest eigenvalue in the GUE Dyson's Brownian motion model.

Next we consider the negative ($x<0$) region. 
We want to study the fluctuation of the scaled height,
\begin{equation}
\label{A1}
 A_1(\tau) = 2 ~t^{-1/3} \left(t/2-h(x=-\frac32 t-t^{2/3}\tau,t)\right)
\end{equation}
as $t\to \infty$.
This corresponds to setting
\begin{align}
 r_j &= t + t^{2/3}\t_j/2, \\
 x_j + 2 r_j -2 &= t/2 - t^{1/3} \xi_j/2,
\end{align}
in the kernel and take the $t\to \infty$ limit.
In the scaling limit, a contribution from the pole at 
$z=-1-w$ in (\ref{Kint}), 
\begin{align}
\label{K1}
 \tilde{K}_1(r_1,x_1;r_2,x_2)
 &=
 \int_{\G_{-1}} \frac{dw}{2\pi i} 
 \frac{(-w)^{x_2+r_1+r_2-2} e^{(2w+1)t}}{(1+w)^{x_1+r_1+r_2-1}},
\end{align}
turns out to be dominant. 
Applying the saddle point analysis to (\ref{K1}), 
we get the limiting kernel,
\begin{equation}
 \K_1(\t_1,\xi_1;\t_2,\xi_2)
 =
 \tilde{\K}_1(\t_1,\xi_1;\t_2,\xi_2)-\Phi_1(\t_1,\xi_1;\t_2,\xi_2)
\end{equation}
where
\begin{align}
 \tilde{\K}_1(\t_1,\xi_1;\t_2,\xi_2)
 &=
 e^{(\t_2-\t_1)(\xi_1+\xi_2)/4+(\t_2-\t_1)^3/12} 
 \Ai\left(\frac{\xi_1+\xi_2}{2}+\frac{(\t_2-\t_1)^2}{4}\right), \\
 \Phi_1(\t_1,\xi_1;\t_2,\xi_2)
 &= 
 \frac{1}{\sqrt{8\pi(\t_2-\t_1)}} 
 \exp\left[-\frac{(\xi_2-\xi_1)^2}{8(\t_2-\t_1)}\right]
\end{align}
for $\t_1<\t_2$ and $\Phi_1(\t_1,\xi_1;\t_2,\xi_2)=0$ for 
$\t_1 \geq \t_2$.
The result is valid for the whole negative ($x<0$) region.
The fact that the multi-point joint distribution of the 1D KPZ
surface on a flat substrate is given by the Fredholm determinant
of this kernel is another main result of this letter.

We remark that the Fredholm determinant of the kernel
with $\t_1=\t_2 (=\t)$,
\begin{equation}
 \K_1(\t,\xi_1;\t,\xi_2) 
 =
 \frac12  \Ai\left(\frac{\xi_1+\xi_2}{2}\right),
\end{equation}
gives the one point height fluctuation in the negative
($x<0$) region and hence should be equivalent to the GOE 
Tracy-Widom distribution function $F_1(s)$ \cite{TW1996}. 
Though we do not know how to prove this directly at the moment, 
the statistics computed from the kernel agrees very well 
with the GOE values numerically. 

The kernels, $\K_1$ and $\K_2$, have a lot of statistical 
information about the surface. 
For instance, using the joint distribution at time $0$ and $\tau$,
we can compute 
\begin{equation}
 g_j(\t) = \sqrt{\langle (A_j(\t)-A_j(0))^2 \rangle/2}
\end{equation}
for $j=1,2$. Here $A_1(\tau)$ is defined in (\ref{A1}) and 
$A_2(\t)$ is the Airy process \cite{PS2002b,Johansson2003}. 
It is clear that $\lim_{\t\downarrow 0}g_j(\t) = 0$
and $g_1(\t)$ (resp. $g_2(\t)$) approaches the standard deviation 
of the GOE (resp. GUE) largest eigenvalue 1.2680 (resp. 0.9018). 
In Fig. 2, we show the comparison of this quantity 
between our theoretical predictions and Monte-Carlo simulation 
results for the surface growth model. The agreement is very good.

Combined with the conjecture in \cite{Ferrari2004}, the fluctuation 
of the largest eigenvalue in the GOE Dyson's Brownian motion model
may also be described by the kernel, $\K_1$. 
It is desirable to have a better understanding of this connection.
It is also an interesting question to see if the joint distribution 
for the GOE case satisfies some differential equations. 
They would be helpful to understand the properties of our process 
$A_1(\t)$. For $A_2(\t)$, such differential equations are already 
known \cite{TW2004,AvM2004}.

To summarize, we have shown that the Green's function 
for the TASEP can be interpreted as a vicious walk problem.
This demonstrates a hidden free fermionic structure behind the 
TASEP and opens up a possibility of analyzing time dependent 
properties of the model in the language of free fermion.
By using the formula, we have found an exact expression for the 
multi-point joint distribution of the KPZ surface for the 
flat initial condition. It is written in the form of the Fredholm 
determinant and the kernel in the scaling limit has been explicitly 
obtained. This is expected to be universal for the KPZ surface
on a flat substrate and may have relevance for the study of the 
GOE Dyson's Brownian motion. 
More detailed analysis and results for the discrete TASEP will 
be reported elsewhere \cite{Sp}.

The author would like to thank H. Spohn, M. Katori, H. Tanemura, 
G. M. Sch\"utz, T. Nagao, M. Pr\"ahofer, A. R{\'a}kos, P. Ferrari
and T. Imamura for useful discussions and comments.
This work is partly supported by the Grant-in-Aid for
Young Scientists (B), the Ministry of Education, Culture, Sports,
Science and Technology, Japan.

\newpage
\begin{large}
\noindent
Figure Captions
\end{large}

\vspace{10mm}
\noindent
Fig. 1:
The initial configuration of the ASEP and the 
corresponding surface. A surface after some time
is also shown with its asymptotic position (dotted line).

\vspace{10mm}
\noindent
Fig. 2:
Comparison of the correlation $g_j(\t)$ ($j=1,2$) computed
from the Fredholm expressions (solid lines) and 
Monte-Carlo simulations (circles). 
The upper ones are for the negative ($x<0$) region and the 
lower ones for the positive ($x>0$) region.

\newpage
\renewcommand{\thepage}{Figure 1}
\begin{picture}(400,300)
\psfrag{h}{}
\psfrag{x}{\hspace{30mm}$x$}
\put(50,100){\includegraphics{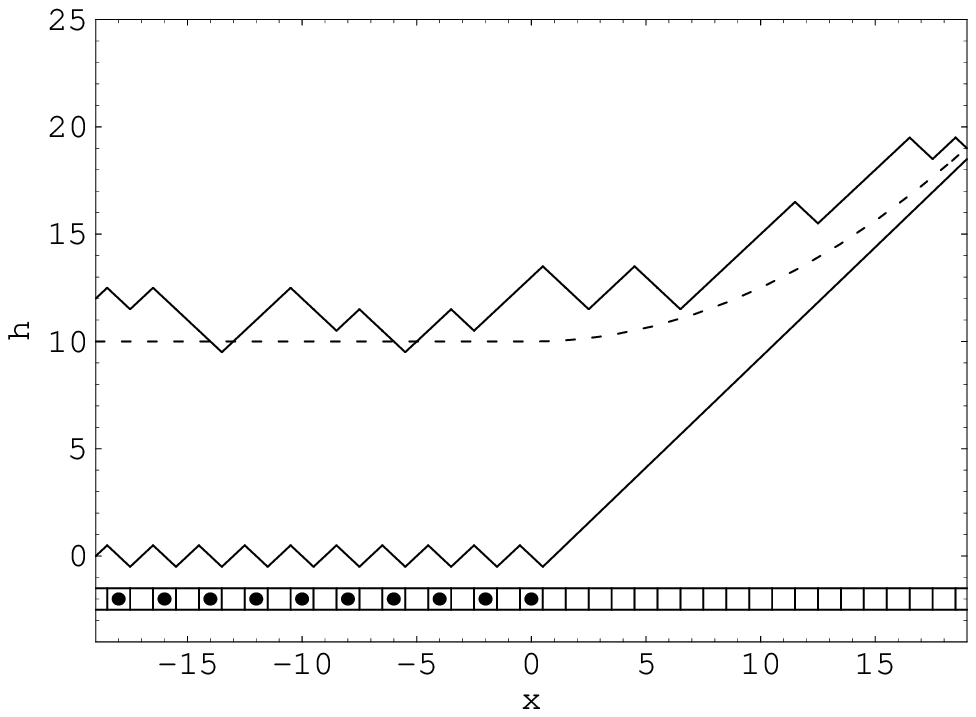}}
\put(40,280){$h$}
\end{picture}

\newpage
\renewcommand{\thepage}{Figure 2}
\begin{picture}(400,300)
\psfrag{Cor}{}
\psfrag{tau}{\hspace{30mm}$\tau$}
\put(50,100){\includegraphics{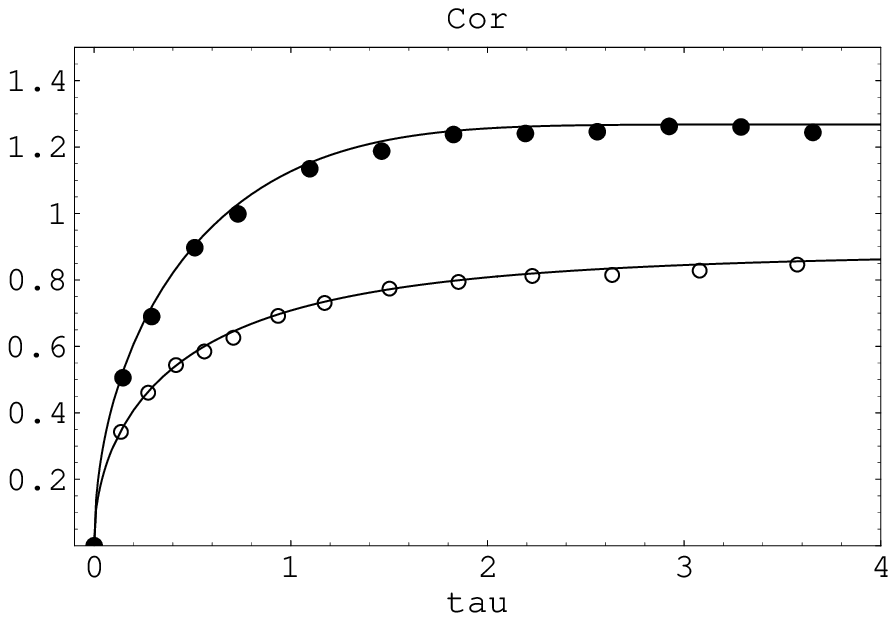}}
\put(40,250){$g_j(\tau)$}
\end{picture}


\begin{thebibliography}{10}
\bibitem{Barabasi1995}
A.-L. Barab\'asi and H. E. Stanley.
\newblock {\em Fractal concepts in surface growth}
\newblock (Cambridge, 1995).

\bibitem{Me1998}
P.~Meakin.
\newblock {\em Fractals, scaling and growth far from equilibrium}
\newblock (Cambridge, 1998).

\bibitem{KPZ1986}
{M. Kardar, G. Parisi and Y. C. Zhang}.
\newblock {\em Phys. Rev. Lett.}, 56:889--892, 1986.

\bibitem{KS1992}
J.~Krug and H.~Spohn.
\newblock In C.~Godr\`eche, editor, {\em Solids far from Equilibrium: Growth,
  Morphology and Defects}, pages 479--582, 1992.

\bibitem{BDJ1999}
{J. Baik, P. A. Deift and K. Johansson}.
\newblock {\em J. Amer. Math. Soc.}, 12:1119--1178, 1999.

\bibitem{PS2000a}
M.~Pr{\"a}hofer and H.~Spohn.
\newblock {\em Phys. Rev. Lett}, 84:4882--4885, 2000.

\bibitem{Jo2000}
K.~Johansson.
\newblock {\em Commun. Math. Phys.}, 209:437--476, 2000.

\bibitem{Mehta2004}
M.~L. Mehta.
\newblock {\em Random Matrices}
\newblock (Elsevier, 3rd edition, 2004).

\bibitem{BR2000}
J.~Baik and E.~M. Rains.
\newblock {\em J. Stat. Phys}, 100:523--541, 2000.

\bibitem{PS2002p}
M.~Pr{\"a}hofer and H.~Spohn.
\newblock {\em J. Stat. Phys.} 115:255-279, 2004.

\bibitem{Dyson1962}
F. J. Dyson.
\newblock {\em J. Math. Phys.}, 3:1191--1198, 1962.

\bibitem{PS2002b}
M.~Pr{\"a}hofer and H.~Spohn.
\newblock {\em J. Stat. Phys.}, 108:1071--1106, 2002.

\bibitem{Johansson2003}
K.~Johansson.
\newblock {\em Com. Math. Phys.} 242:277-329, 2003.

\bibitem{Johansson2005}
K.~Johansson.
\newblock {\em Ann. Prob.} 33:1-30, 2005.

\bibitem{SI2004}
T.~Sasamoto and T.~Imamura.
\newblock {\em J. Stat. Phys.}, 115:749-803, 2004.

\bibitem{IS2004b}
T. Imamura and T. Sasamoto.
\newblock{Nucl. Phys. B}, 699: 503--544, 2004.

\bibitem{BR2004p}
A. Borodin and E. M. Rains.
math-ph/0409059.

\bibitem{Calogero1969}
F. Calogero.
\newblock {\em J. Math. Phys.}, 10:2197--2200, 1969.

\bibitem{Sutherland1972}
B. Sutherland.
\newblock {\em Phys. Rev. A}, 5:1372--1376, 1972.

\bibitem{Ferrari2004}
P. Ferrari.
\newblock {\em Commun. Math. Phys.}, 252:77--109, 2004.

\bibitem{Liggett1985}
T.~M. Liggett.
\newblock {\em Interacting Particle Systems}
\newblock (Springer-Verlag, 1985).

\bibitem{Liggett1999}
T.~M. Liggett.
\newblock {\em Stochastic Interacting Systems: Contact, Voter, and Exclusion
  Processes}
\newblock (Springer-Verlag, 1999).

\bibitem{Spohn1991}
H.~Spohn.
\newblock {\em Large Scale Dynamics of Interacting Particles}
\newblock (Springer-Verlag, 1991).

\bibitem{Schuetz2000}
G.~M. Sch{\"u}tz.
\newblock Exactly solvable models for many-body systems far from equilibrium,
\newblock in: C.~Domb and J.~L. Lebowitz (editors)
{\em Phase Transitions and Critical Phenomena 19} (2000).

\bibitem{GS1992}
{L.-H. Gwa and H. Spohn}.
\newblock {\em Phys. Rev. Lett.}, 68:725--728, 1992.

\bibitem{DL1998}
{B. Derrida and J. L. Lebowitz}.
\newblock {\em Phys. Rev. Lett.}, 80:209--213, 1998.

\bibitem{Schuetz1997}
G.~M. Sch{\"u}tz.
\newblock {J. Stat. Phys.} 88:427--445, 1997.

\bibitem{NS2004}
T.~Nagao and T.~Sasamoto.
\newblock{Nucl. Phys. B}, 699: 487--502, 2004.

\bibitem{RakosSchuetz2004p}
{A. R\'akos and G. M. Sch\"utz}.
cond-mat/0405464.

\bibitem{Fisher1984}
M. E. Fisher.
\newblock {\em J. Stat. Phys.}, 34:667--729, 1984.

\bibitem{Forrester1989}
P. J. Forrester.
\newblock {\em J. Stat. Phys.}, 56:767--782, 1989.

\bibitem{NF1998}
T.~Nagao and P. J. Forrester.
\newblock{Phys. Lett. A}, 247:42--46, 1998.

\bibitem{TW1996}
C.~A. Tracy and H.~Widom.
\newblock {\em Commun. Math. Phys.}, 177:727--754, 1996.

\bibitem{TW2004}
C.~A. Tracy and H.~Widom.
\newblock {\em Elect. Comm. in Probab.}, 8:93--98, 2003;
\newblock {\em Commun. Math. Phys.}, 252:7--41, 2004.

\bibitem{AvM2004}
M. Adler and P. van Moerbeke.
math.PR/0302329;
math.PR/0403504.

\bibitem{Sp}
T. Sasamoto.
in preparation.

\end{thebibliography}
\end{document}